\documentstyle[epsf]{mn}
\title{Kinematics and Metallicities of Globular Clusters in M104}

\author[T.J. Bridges et al.]
{T.J. Bridges$^{1,6}$, K.M. Ashman$^2$, 
S.E. Zepf$^3$\thanks{Visiting Astronomer,
William Herschel Telescope.
The WHT is operated on the island of La Palma by the Royal Greenwich
Observatory at the Spanish Observatorio del Roque de los Muchachos of
the Instituto de Astrofisica de Canarias.}
\thanks{Hubble Fellow},
D. Carter$^{1 *}$, 
\newauthor
D.A. Hanes$^{4  *}$, 
R.M. Sharples$^{5  *}$, and
J.J. Kavelaars$^{4  *}$ \\
$^1$Royal Greenwich Observatory,
Madingley Road,
Cambridge, England,
CB3 0EZ\\
$^2$Dept. of Physics \& Astronomy, University of Kansas,
Lawrence, KS, USA \\
$^3$Astronomy Dept, Univ. of California, Berkeley, CA, USA \\
$^4$Physics Dept, Queen's University, Kingston, Ontario, Canada \\
$^5$Dept. of Physics, University of Durham, Durham \\
$^6$E-mail:~~tjb@ast.cam.ac.uk}

\date{Accepted 20 August 1996.   Received 23 May 1996.}

\pubyear{1996}

\begin{document}

\maketitle

\begin{abstract}

We have obtained spectra for globular cluster candidates in M104
with LDSS-2 on the William Herschel Telescope, confirming 34 objects
as M104 globular clusters.
We find a cluster velocity
dispersion of $\sim$ 260 km/sec, and the Projected Mass Estimator gives
a mass of 5.0 (3.5,6.7) $\times$ 10$^{11}$ M$_\odot$ for M104
within a projected radius of $\sim$ 330$^{\prime \prime}$ 
(14 kpc for D=8.55 Mpc).  Our best estimate for 
the mass-to-light ratio
is M/L$_{V_T}$= 16$^{+5.5}_{-5.0}$ within the same radius.  Considering
all of the possible sources of uncertainty, we find a lower limit 
of M/L$_V$= 5.3, which is larger than the M/L$_V$ found from rotation
curve analyses inside 180$^{\prime \prime}$.  We thus conclude that
the mass-to-light ratio increases with radius, or in other words  
that M104 possesses a dark matter halo.   There is a 
marginal detection of rotation in the M104 cluster system at
the 92.5\% confidence level; larger samples
will be needed to investigate this possibility.  Interestingly,
the M104 globular cluster and planetary nebulae (PNe)
kinematics are roughly consistent inside $\sim$ 
100$^{\prime \prime}$. 
Finally, we find a mean
cluster metallicity of [Fe/H] = $-$0.70 $\pm$ 0.3, which is more
typical of clusters in gE/cD galaxies than it is of clusters in
other spirals. 

\end{abstract}


\section{Introduction}

Globular cluster spectra provide an excellent means to study the 
kinematics and metallicities of old stellar populations, and to
probe the mass distributions of their parent galaxies (see Brodie
1993 and Zepf 1995 for recent reviews).
Globular clusters and planetary nebulae
(PNe) bridge the gap between the inner regions of ellipticals
and early-type spirals, where
masses and M/L ratios can be determined from rotation curve work
or integrated light techniques, and the outermost regions which can
be studied via X-ray emission from hot gas. 

In our Galaxy, it is well-known that two globular cluster populations
exist--a rapidly rotating, metal-rich disk, and a slowly rotating
and metal-poor halo (Zinn 1985; 
Armandroff \& Zinn 1988; Armandroff 1989).  
The same situation appears to hold 
in M31, though this is more controversial (Huchra 1993;
Ashman \& Bird 1993).  In M33, there is no
significant rotation in the old, metal-poor clusters (Schommer et al.
1991).
To date, M81
is the only late-type galaxy outside the Local Group with measured
cluster velocities.  Perelmuter, Brodie \& Huchra (1995) obtained
velocities for 25 clusters out to 20 kpc from the galaxy center, and
found a mass of $\sim$ 3 $\times$ 10$^{11}$ M$_\odot$ for M81 within
the same radius using the Projected Mass Estimator (PME).  Their data
``cannot be used to demonstrate rotation" in the cluster system,
though the cluster velocities are consistent with the HI rotation
curve.

There exist a handful of ellipticals with measured cluster velocities:
NGC 5128 (87: Harris, Harris \& Hesser 1988), NGC 1399 (47: Grillmair et al.
1994), M87 (44), and M49 (26) (Mould et al. 1987, 1990).  For
M87 and M49, the cluster system has a rotation of $\sim$ 200 km/sec 
(for {\it all} clusters combined--no separation has been done by
colour/metallicity).  There is {\it no} rotation seen in the metal-poor
clusters in NGC 5128 and NGC 1399, while there is significant rotation
seen in their PNe at a level of 100--300 km/sec (Hui et al. 1995,
Arnaboldi et al. 1994).
In NGC 5128, the {\it metal-rich}
cluster population identified by Zepf \& Ashman (1993) is observed
to rotate like the PNe.
Such
kinematical differences between clusters and PNe are extremely 
provocative, and we will return to this 
point later.

M104 (NGC 4594; the Sombrero) has been well-studied spectroscopically
in its inner regions (see Table 1 for basic properties
of M104).  Faber et al (1977) and Schweizer (1978) 
derived major axis rotation curves from ionized gas, stellar absorption
lines and 21cm emission. Schweizer found a rotation velocity of 
$\sim$ 300--350 km/sec between 5--15 kpc, and a total mass of 
3.3 $\times$ 10$^{11}$ M$_\odot$  and a M/L$_B$ of
4.9 $\pm$ 1.2 within 15 kpc.
These values are all in reasonable agreement with those found by
Faber et al.  Kormendy \& Illingworth (1982) obtained the velocity
and velocity
dispersion profiles along the M104 major axis from optical absorption
lines, finding that V flattens out at $\sim$ 250 km/sec between
40--120$^{\prime \prime}$ from the galaxy center, while $\sigma$ flattens out
to $\sim$ 100 km/sec at the same distance.  These values are in good
agreement with those found by Kormendy (1988), Jarvis \& Dubath
(1988), and Hes \& Peletier (1993), although these authors were more
interested in the behavior very near the galaxy center.  

Most authors
have not found any rotation around the minor axis, with  the exception
of Hes \& Peletier who did find some minor axis rotation.  Kormendy
\& Westpfahl (1989) showed that 2 $\leq$ M/L$_V$ $\leq$ 4 between
0.5 $\leq$ r $\leq$ 180$^{\prime \prime}$. 
Hes \& Peletier found absorption-line strength gradients along the
major axis, and concluded on the basis of these gradients and the
central kinematics that the M104 bulge is similar in many respects to
a giant elliptical.  Jarvis \& Freeman (1985) constructed models which
showed that the surface brightness profiles and kinematics are 
consistent with the M104 bulge being an isotropic oblate spheroid,
flattened mostly by rotation, {\it unlike} the case for ellipticals
of the same luminosity.  Jarvis \& Freeman found a M/L$_V$ = 3.6
within $\sim$ 80$^{\prime \prime}$, and 
a disk/bulge ratio of $\sim$ 0.25 for
M104.  Finally, Burkhead (1986) has carried out a major 
photometric study and decomposition of M104.

Previous photometric studies of the M104 globular clusters include 
Wakamatsu (1977), Harris et al. (1984), and Bridges \& Hanes (1992).
The last two studies found a cluster specific frequency S$_N$ of 2--3,
while Bridges et al. estimated the clusters to have [Fe/H] $\simeq$
$-$0.8 from B$-$V colours, a metallicity higher than that found for
the clusters in the Milky Way and other nearby spirals.  

Globular cluster candidates for spectroscopic
analysis were selected in two ways.  First, the COSMOS team carried out
astrometry and crude photometry on `V' 
(AAT 1853, IIaD + GG485) and  `B' (AAT 1859, IIIaJ + GG385)
AAT plates,  giving internal positions for thousands of objects.  The
HST Guide Star Catalogue was then used to identify a set of several
dozen stars which were picked by COSMOS, allowing a transformation 
to absolute coordinates to better than 0.2$^{\prime \prime}$ rms.
COSMOS had trouble detecting and measuring objects close to the galaxy
bulge, but we also had B,V CCD photometry covering three small patches
of the M104 bulge (Bridges \& Hanes 1992).  These photometrically
calibrated data then allowed us to crudely calibrate the COSMOS data
by magnitude and colour, and to establish the astrometric scales in
the CCD data so that we could add a few near-central objects to our
target lists.  Final object selection was made on the basis of 
stellar appearance, and by cuts in magnitude and colour.
%!!HOW EXACTLY??!!
In all, 103 candidates in
3 fields were obtained, extending in radius between
0.8 to 6.2 arcmin from the
galaxy center. 
%!!WHAT MAGNITUDE CUTS?!!
The LEXT package was used to produce input x, y
files which were then used to produce the punched masks. A slit width
of 1.5$^{\prime \prime}$ 
was used, and the slit length varied between 10--60 arcsec.

\section{Data}

\subsection{Observations and Data Reduction}

Spectra for 76 cluster candidates in two of these fields 
were obtained with the Low Dispersion
Survey Spectrograph (LDSS-2) in April 1994;
see Table 2 for further details of the observing.
Dome and twilight flats were taken
at the beginning and end of each night, and CuAr arcs were taken
throughout each night.  Finally, long-slit spectra of M13, M92,
NGC 6356, and radial velocity standard stars were taken for velocity
and metallicity calibration.
Except where
indicated, the LEXT package was used
for all data reduction.  First, 
the MAKEFF task was used to produce a domeflat of mean unity, which
was then divided into all of the program frames.  After checking to
ensure that there were no spatial or spectral shifts between the 
program frames, the LCCDSTACK program (kindly supplied by Karl
Glazebrook) was used to combine the program frames for each mask in 
an optimal way; all cosmic rays were removed during the stacking.
Wavelength calibration was then done using the ARC task on the CuAr
exposures; a 3rd order polynomial was found to give a satisfactory
fit with residuals typically $\leq$ 0.1 $\AA$.  Finally, the spectra
were optimally extracted and sky subtracted, generally
with linear fits for the background sky.  71 spectra were extracted
in total, though many of the spectra are of low S/N.  
Figure 1 
shows representative spectra ranging from low to high S/N for three
confirmed globular clusters.

\begin{table}
\caption{Basic and Derived Information about M104.}  
\begin{tabular}{ccc} 
\hline\hline
Quantity & Value & Reference \\
\hline
%  & &  \\
 RA (1950) & 12 37 22.80 & 1  \\
 Dec (1950) & $-$11 21 00.0 & 1 \\
 V$_{hel}$ & 1091 km/sec & 1 \\
 Hubble Type & Sa$^+$/Sb$^-$ & 2 \\
 M$_B$ & $-$20.4 & 3 \\
 Adopted Distance & 8.55 Mpc & 4 \\
 Specific Frequency & 2 $\pm$ 1 & 5 \\
 Cluster [Fe/H] & $-$0.7 $\pm$ 0.3 & 6 \\
 M/L$_{B_T}$ &  22$^{+7.5}_{-6.5}$ & 6 \\
\hline
\end{tabular}
\bigskip

\noindent References

\bigskip

(1)~~NASA/IPAC Extraglactic Database (NED) \\
(2)~~Sandage \& Tammann (1981) \\
(3)~~Burkhead (1986), taking D=8.55 Mpc \\
(4)~~Ciardullo, Jacoby, \& Tonry (1993) \\
(5)~~Bridges \& Hanes (1992) \\
(6)~~This work

\end{table}

\begin{table}
\caption{Observing Log}
\begin{tabular}{cc} 
\hline\hline
 Dates & April 11-14 1994 \\
 Telescope/Instrument & 4.2m WHT/LDSS-2 \\
 Dispersion (Resolution) & 2.4 $\AA$/pixel (6 $\AA$ FWHM) \\
 Detector & 1024$^2$ TEK CCD \\
 Wavelength Coverage & 3800--5000 $\AA$ \\
 Seeing & 1--2$^{\prime \prime}$ \\
 Exposure time (Mask \#1/\#2) & 4.5/2.5 hr \\
 Mean Airmass (Mask \#1/\#2) & 1.4/1.5 \\
 Number Objects (Mask \#1/\#2) & 44/32 \\
\hline
\end{tabular}
\end{table}

\subsection{Radial Velocities and Confirmed Globular Clusters}

The extracted spectra were first scrunched onto a log wavelength scale
and then cross-correlated with template spectra of M13, M92, NGC 6356
and HD172, using the IRAF FXCOR task.  By experimentation, we found
that cross-correlations with peak heights less than 0.1 were not reliable,
and hence were not used.  We further demanded that bona-fide globular
clusters have two or more reliable cross-correlations. 
Table 3 shows our final velocities for the 71 extracted spectra
in the two fields.  The velocity shown in Table 3 is the
mean (weighted by the cross-correlation peak height) for the four
templates;  `??' means that no spectrum was extracted and `**'
signifies that 
no reliable cross-correlation could be obtained.

\begin{table}
\caption{Velocities of 
Globular Cluster Candidates in M104.  Successive columns
give Id \#, Ra, Dec, Velocity, and Velocity Error, for 44
cluster candidates in Field \#1.  A `??' means that no
spectrum could be extracted, and a `**' means that no reliable
cross-correlation could be obtained.  There are 34 confirmed
globular clusters (see last column).}
\begin{tabular}{lllllc} 
\hline\hline
Id & Ra  & Dec  & V$_{hel}$ & Error \\
  & (1950) & (1950) & (km/s) & (km/s) & Cluster? \\
\hline
 1$-$1 & 12 37 40.460 &  -11 21 33.55 &   212  &  18  & N \\
 1$-$2 & 12 37 34.548 &  -11 20 18.29 &   776  &  11  & Y \\
 1$-$3 & 12 37 35.516 &  -11 21 47.40 &  1369  & 129  & Y \\
 1$-$4 & 12 37 33.716 &  -11 18 53.53 &  1152  &  40  & Y \\
 1$-$5 & 12 37 34.864 &  -11 19 27.65 &   755  &  61  & Y\\
 1$-$6 & 12 37 26.784 &  -11 18 10.02 &   109  &  20  & N \\
 1$-$7 & 12 37 28.576 &  -11 19  8.93 &   **  & **  & N \\
 1$-$8 & 12 37 27.544 &  -11 16 14.52 &   **  & ** & N \\
 1$-$9 & 12 37 28.896 &  -11 17 36.80 &   194  &  29 & N \\
 1$-$10 & 12 37 36.140 &  -11 24 20.96 &   -51  &  21 & N \\
 1$-$11 & 12 37 28.848 &  -11 19 54.47 &  1457  &  12 & Y \\
 1$-$12 & 12 37 37.988 &  -11 22 11.38 &  1370  &  32 & Y \\
 1$-$13 & 12 37 36.624 &  -11 18 30.11 &   219  &  78 & N \\
 1$-$14 & 12 37 33.112 &  -11 17  0.81 &   **  & ** & N \\
 1$-$15 & 12 37 35.600 &  -11 25 29.50 &  1035  &  24 & Y \\
 1$-$16 & 12 37 39.288 &  -11 21  7.67 &  1256  &  18 & Y \\
 1$-$17 & 12 37 35.336 &  -11 22 26.42 &   **  & ** & N \\
 1$-$18 & 12 37 12.224 &  -11 24 59.35 &   **  & ** & N \\
 1$-$19 & 12 37 20.336 &  -11 21 58.89 &   573  &  21 & Y \\
 1$-$20 & 12 37  5.104 &  -11 19 40.84 &  1186  &  51 & Y \\
 1$-$21 & 12 37 35.580 &  -11 16 37.59 &   **  & ** & N \\
 1$-$22 & 12 37 31.244 &  -11 22 56.33 &   131  &  43 & N \\
 1$-$23 & 12 37 41.120 &  -11 25  3.76 &   **  & ** & N \\
 1$-$24 & 12 37 24.920 &  -11 23 12.60 &    ??  &   ?? & N \\
 1$-$25 & 12 37  7.988 &  -11 21  2.23 &  1231  &  26 & Y \\
 1$-$26 & 12 37 28.316 &  -11 23 51.23 &  1199  &  14 & Y \\
 1$-$27 & 12 37  3.460 &  -11 24 23.50 &  -233  &  26 & N \\
 1$-$28 & 12 37 18.652 &  -11 24  8.86 &   932  &  25 & Y \\
 1$-$29 & 12 37 37.460 &  -11 22 45.05 &   853  &  31 & Y \\
 1$-$30 & 12 37 17.232 &  -11 23 25.00 &   **  & ** & N \\
 1$-$31 & 12 37  6.112 &  -11 16 49.60 &   **  & ** & N \\
 1$-$32 & 12 37 14.744 &  -11 22 27.14 &  1025  & 203 & Y \\
 1$-$33 & 12 37 25.420 &  -11 20  5.94 &   448  & 141 & N \\
 1$-$34 & 12 37 11.292 &  -11 18 24.53 &   -31  &   6 & N \\
 1$-$35 & 12 37 16.768 &  -11 22 43.71 &  1300  &  40 & Y \\
 1$-$36 & 12 37 35.436 &  -11 26 11.98 &   **  & ** & N \\
 1$-$37 & 12 37 14.296 &  -11 16 37.64 &  2456  &  24 & N \\
 1$-$38 & 12 37 18.800 &  -11 24 42.64 &   **  & ** & N \\
 1$-$39 & 12 37  6.072 &  -11 18 55.96 &   832  &  61 & Y \\
 1$-$40 & 12 37 26.668 &  -11 25 16.04 &   **  & ** & N \\
 1$-$41 & 12 37  7.676 &  -11 20 24.56 &   **  & ** & N \\
 1$-$42 & 12 37  3.752 &  -11 23 11.61 &   **  & ** & N \\
 1$-$43 & 12 37 16.964 &  -11 22 14.78 &   **  & ** & N \\
 1$-$44 & 12 37 41.244 &  -11 20 52.17 &   **  & ** & N \\
\hline
\end{tabular}
\end{table}

\begin{table}
\contcaption{Id \#, Ra, Dec, Velocity, and
Velocity Errors are given for 32 cluster candidates in Field \#2.}
\begin{tabular}{lllllc} 
\hline\hline
Id & Ra  & Dec  & V$_{hel}$ & Error \\
  & (1950) & (1950) & (km/s) & (km/s) & Cluster? \\
 2$-$1 & 12 37 40.460 &  -11 21 33.55 &   306  &  45 & N \\
 2$-$2 & 12 37 30.192 &  -11 20 20.00 &   808  &  39 & Y \\
 2$-$3 & 12 37 36.332 &  -11 21 45.12 &  1524  &  14 & Y \\
 2$-$4 & 12 37 16.272 &  -11 19 13.20 &   968  &  78 & Y \\
 2$-$5 & 12 37 31.388 &  -11 19 24.32 &  1220  &  89 & Y \\
 2$-$6 & 12 37 24.600 &  -11 18 52.32 &  1283  &  45 & Y \\
 2$-$7 & 12 37 17.976 &  -11 19 37.71 &   976  &  24 & Y \\
 2$-$8 & 12 37 23.200 &  -11 22  0.79 &   979  &  14 & Y \\
 2$-$9 & 12 37 39.308 &  -11 25 25.85 &   -83  &  25 & N \\
 2$-$10 & 12 37 23.312 &  -11 20  4.55 &   **  & ** & N \\
 2$-$11 & 12 37 32.396 &  -11 17 30.21 &  1411  &  56 & Y \\
 2$-$12 & 12 37 13.260 &  -11 18  7.37 &   857  &  42 & Y \\
 2$-$13 & 12 37 23.720 &  -11 18 27.38 &   616  &  34 & Y \\
 2$-$14 & 12 37 25.464 &  -11 22 36.75 &  1045  &   7 & Y \\
 2$-$15 & 12 37 26.900 &  -11 22 18.79 &   828  &  71 & Y \\
 2$-$16 & 12 37 16.784 &  -11 19 50.53 &   875  &  22 & Y \\
 2$-$17 & 12 37 35.248 &  -11 23 16.06 &  1275  &  100 & Y \\
 2$-$18 & 12 37  7.460 &  -11 23  3.04 &   **  & ** & N \\
 2$-$19 & 12 37 34.280 &  -11 16 18.73 &     ??  &   ?? & N \\
 2$-$20 & 12 37 37.480 &  -11 22 51.50 &   **  & ** & N \\
 2$-$21 & 12 37 34.952 &  -11 24 25.56 &   **  & ** & N \\
 2$-$22 & 12 37  5.504 &  -11 23 17.64 &     ??  &   ?? & N \\
 2$-$23 & 12 37  6.072 &  -11 24 50.98 &     ??  &   ?? & N \\
 2$-$24 & 12 37 30.260 &  -11 23 46.25 &  1505  & 119 & Y \\
 2$-$25 & 12 37  7.380 &  -11 21  3.77 &   **  & ** & N \\
 2$-$26 & 12 37 40.516 &  -11 16 50.78 &     ??  &   ?? & N \\
 2$-$27 & 12 37 13.332 &  -11 16 53.25 &   **  & ** & N \\
 2$-$28 & 12 37 43.736 &  -11 23 31.64 &   **  & ** & N \\
 2$-$29 & 12 37 21.284 &  -11 16 34.38 &   **  & ** & N \\
 2$-$30 & 12 37  5.916 &  -11 24 31.71 &  1104  & 104 & Y \\
 2$-$31 & 12 37 15.416 &  -11 21 48.77 &   939  &  21 & Y \\
 2$-$32 & 12 37 12.236 &  -11 24  9.03 &   **  & ** & N \\
\hline
\end{tabular}
\end{table}

A word on velocity {\it uncertainties} is in order.  The errors given
in Table 3 are merely the rms amongst the four templates.  We do not
have a good idea of the {\it external} uncertainties.  There is one 
object in common between the two masks (\# 1--1~=~2--1), with a velocity
difference of 95 km/sec--however, it is a foreground object
(we were not able to observe our third field in M104, which
had several objects in common with the other two, because of
the higher priority placed on NGC 4472, our principal
target).  For lack
of information, we assume velocity errors of 50--100 km/sec, values
found from other data of comparable S/N (e.g. M81: Perelmuter, Brodie,
\& Huchra 1995; M87 \& M49: Mould et al. 1990).  In any event, even
velocity errors of 100 km/sec have little effect on our mass and M/L
determinations, given that the observed 
velocity dispersion of our confirmed
cluster sample is $\sim$ 250 km/sec (Section 3.1 below).

In order to isolate a sample of true globular clusters, we subjected
the velocity distribution of the (46) objects in Table 3 with 
reliable velocities (see Figure 2)
to the KMM mixture modelling analysis (McLachlan \&
Basford 1988; Ashman, Bird, \& Zepf 1994).  This analysis found a best
fit of 3 groups--one high velocity point (\#1--37), a foreground group 
of 11 objects, and 34 objects with velocities 500 $\leq$ V $\leq$ 1600 
km/sec.  Inspection of Figure 2 suggests that this is a sensible 
split.  The last column of Table 3 shows which objects are confirmed
globular clusters; there are 34 such objects in total.

\section{Results}

\subsection{Mass and M/L Ratio}

We have used the ROSTAT code (Beers, Flynn, \&
Gebhardt 1990; Bird \& Beers 1993), to obtain robust values for the
mean cluster velocity and velocity dispersion.  We find that the 
robust
mean velocity (location) is 1074 ($-$79,$+$78) km/sec, where we quote
boot-strapped 90\% confidence intervals.  Similarly, the 
biweight velocity
dispersion (scale) is 255 (210, 295) km/sec, where this value has been
corrected for a possible rotation of $\sim$ 70 km/sec (Section 3.2.1), 
and velocity errors of 100 km/sec per point are assumed;
we again quote 90\% boot-strapped confidence intervals.
The correction for possible rotation
has been done assuming a step function
(cf. Mould et al. 1990, and see Section 3.2.1 and Figure 3); this 
correction changes
$\sigma$ by only $\sim$ 10 km/sec. 
The mean velocity
is reassuringly close to the recession velocity of M104 itself, which
is 1091 $\pm$ 5 km/sec (RC3). 
The M104 cluster velocity dispersion is much larger than
that of old clusters in
other spiral galaxies (e.g. Milky Way: $\sigma$ $\simeq$ 
100 km/sec--Armandroff 1989, Da Costa \& Armandroff
1995; M31: $\sigma$ $\simeq$ 150 
km/sec--Huchra 1993; M33: $\sigma$ $\simeq$ 70 km/sec--Schommer 
et al. 1991); 
M81: $\sigma$ $\simeq$ 150 km/sec--Perelmuter, Brodie \& Huchra 1995),
but smaller than that of gE galaxies (e.g. M87: $\sigma$ $\simeq$ 
385 km/sec; M49:~ $\sigma$ $\simeq$ 330 km/sec--Mould et al. 1990;
NGC 1399:~$\sigma$ $\simeq$ 390 km/sec--Grillmair et al. 1994).
While a direct comparison of these numbers
isn't very meaningful, since they are measured to different radii
(or, more physically, scale-lengths) in each galaxy, they
do indicate the presence of dark matter halos in these galaxies.
In an attempt to see if the cluster velocity dispersion varies with
galactocentric radius, we divided the data into 4 radial bins,
with roughly equal numbers of clusters in each bin.  Unfortunately,
the small number of data points and resulting large confidence intervals
mean that we cannot place useful constraints on any such variation.

We have used the Projected Mass Estimator (PME) to determine the mass
of M104:

\[M_p={f_p\over NG} \sum_{i=1}^{N} r_{i} v_{i}^{2} \]

\noindent where r$_i$ is the projected galactocentric radius of the
{\it i}th cluster, and v$_i$ is the velocity corrected for the mean
cluster velocity (1074 km/sec).  The f$_p$ factor depends on the
assumed cluster velocity distribution.  We have adopted a value of
f$_p$ = 16/$\pi$ which assumes isotropic orbits and a central point 
mass (Bahcall \& Tremaine 1981; Heisler, Tremaine, \&
Bahcall 1985); for radial and tangential orbits
f$_p$ = 32/$\pi$ and 32/3$\pi$ respectively. 
Finally, we have assumed a distance D= 8.55 Mpc based on the SBF distance
of Ciardullo, Jacoby, \& Tonry (1993; Ford et al. 1996 give a slightly
higher distance of 8.9 $\pm$ 0.6 Mpc, based on the PNLF) for M104--the
mass will scale with D.
The robust implementation of the PME gives
M$_p$= 5.2 (3.9, 6.7) $\times$ 10$^{11}$ M$_\odot$, where we quote 90\%
boot-strapped confidence intervals, and we have taken the galaxy 
centroid as the dynamical center.  Correcting for velocity errors,
possible rotation, and using the clusters themselves to define the
dynamical center reduce M$_p$ by $<$ 10\%.  Thus, we quote
M$_p$= 5.0 (3.5, 6.7) $\times$ 10$^{11}$ M$_\odot$.  The main uncertainties
in M$_p$ are the assumptions about the cluster orbital distribution,
the M104 distance, and
whether or not we use a central point mass or extended mass distribution
for M104, all of which are uncertain by roughly a factor of two.  We take
M$_p$ to be the mass within the projected radius of the furthermost
cluster in our sample, which lies at 5.5 arcmin (14 kpc for 
D=8.55 Mpc). 

From Burkhead (1986), the total integrated magnitude of M104 is
B$_{tot}$ = 9.24 within 5.5 arcmin, where we have corrected for
A$_B$= 0.12 mag (Burstein \& Heiles 1984).  
Thus, L$_B$ ($<$~5.5$^{\prime}$) = 
2.3 $\times$ 10$^{10}$ L$_\odot$ for D=8.55 Mpc, and 
M/L$_{B_T}$ = 22$^{+7.5}_{-6.5}$, where the uncertainties reflect only the
{\it formal} confidence intervals for M$_p$ above.  Taking 
B$-$V = 1 (Burkhead 1986), the corresponding M/L$_V$ = 
16$^{+5.5}_{-5.0}$.  These M/L ratios scale as ${8.55 \rm Mpc} \over {D}$. 
Given the uncertainties mentioned above, our quoted M/L ratios are
probably only believable to within a factor of two.  

Do our data imply the existence of dark matter in the M104 halo?
Kormendy \& Westpfahl (1989) showed that spectroscopic data available
at that time yielded 2 $\leq$ M/L$_V$ $\leq$ 4 between
0.5 -- 180$^{\prime \prime}$, assuming a distance
of 18 Mpc, and they concluded that there
{\it ``is no evidence for halo dark matter between
11 $\leq$ r $\leq$ 215$^{\prime \prime}$."} 
From our cluster data,
we can obtain a lower bound on the M/L ratio
by assuming 
tangential orbits, a central point mass, and a 
distance of 18 Mpc:~ we find that M/L$_V$ 
$\geq$ 5.3 within 5.5 arcmin.  
Thus, we find that the  
M/L ratio must increase with radius (by a factor of $\sim$ 4 between
180--330$^{\prime \prime}$ for our best estimate of M/L$_V$ = 16),
and we conclude that 
{\it there is indeed dark matter in the M104 halo}.

Our M/L agrees well with those found from globular clusters in 
other spirals (e.g. M31: M/L$_B$ = 16--Huchra 1993; M81: M/L$_B$= 19--
Perelmuter, Brodie \& Huchra 1995, both within 20 kpc).  For
some gE/cD galaxies, M/L is larger (cf. M87: M/L$_V$ = 31
inside $\sim$ 40 kpc; NGC 1399: M/L$_B$ $\simeq$ 70--80
between 20--40 kpc.   For
other ellipticals, however, the M/L is lower (e.g. M49: 
M/L$_B$ $\leq$ 10 inside 20 kpc; NGC 5128: M/L$_B$ $\simeq$ 10).

\subsection{Kinematics and Comparison with PNe}

\subsubsection{Rotation in the Cluster System?}

In Figure 3, we show the cluster velocities as a function of
distance along the major axis of M104; as discussed in Section 2.2,
the error bars are only internal, and the true uncertainties are
likely to be 50--100 km/sec.  While there is considerable scatter
at all radii, there is a {\it hint} of rotation in the cluster
system, with clusters west of the galaxy center having lower
velocities on average than those on the eastern side.  In Figure 3,
the solid line is a linear least squares fit, with a slope of 
0.43 km/sec/arcsec. 
Over the
8 arcmin along the major axis covered by the studied clusters,
this amounts to $\sim$ 210 km/sec or a V$_{rot}$ of $\sim$ 
105 km/sec.  
Another way to estimate the possible rotation is to compare the
mean cluster velocity east and west of the minor axis.  We have
done this in a robust way with ROSTAT, since this should be more
reliable than a classical Gaussian estimator with the small number
of datapoints.  We find that the mean velocities for the two
datasets are: V$_{east}$= 1137 ($-$116, $+$124); V$_{west}$ = 991
($-$66, $+$107) km/sec, where the values in brackets are the 90\%
bootstrapped uncertainties.  Therefore, 2V$_{rot}$= 
146$^{+164}_{-133}$ km/sec, or V$_{rot}$= 73$^{+82}_{-66}$ km/sec.

We have also carried out a Mann-Whitney U test on our data.  The
hypothesis that the velocities of clusters with RA less than that 
of M104 are drawn from the same population as the velocities of
clusters with RA greater than M104 can be rejected at the 92.5\%
confidence level.  In addition, the hypothesis that the velocities
of clusters with Dec less than that of M104 are drawn from the same
population as the velocities of clusters with Dec greater than M104
cannot be rejected.  In other words, there is a correlation
between velocity and major axis position (since the M104 major axis
lies almost exactly East-West), and no correlation between velocity
and minor axis position.  We interpret this as a marginal
detection of rotation in the cluster system, at the 92.5\%
confidence level. 
However, it is clear from 
Figure 3 that we have large error bars, and
there may be considerable real scatter;
many more velocities will be needed to resolve this issue.

Rotation at a similar amplitude 
is observed in the outer, metal-rich M31 clusters (V$_{rot}$ $\sim$
70 km/sec: Huchra 1993).  Rotation at lower levels of
40--50 km/sec are found for the metal-rich clusters in M33
(Schommer et al. 1991) and NGC 5128 (Hui et al. 1995), and for the
metal-poor Galactic halo and M31 
clusters (Da Costa \& Armandroff 1995; Huchra 1993), while 
no significant rotation is 
detected in the metal-poor M33 clusters (Schommer et al. 1991). 
It would be extremely interesting to obtain a large sample
of high S/N spectra of M104 clusters, so that a similar
comparison could be made between the kinematics of metal-poor
and metal-rich clusters in that galaxy.

\subsubsection{Comparison with M104 PNe}

Ken Freeman has very kindly shared preliminary results for 100 PNe
velocities in M104 (of $\sim$ 250 total).  For those objects
within 74$^{\prime \prime}$  of the galactic plane, the PNe show a rotation
of 50--100 km/sec out to $\sim$ 250$^{\prime \prime}$ along the major axis.
The PNe velocity dispersion is $\sim$ 220 km/sec at 50$^{\prime \prime}$ 
radius, dropping off to $\sim$ 180 km/sec between 120--150$^{\prime \prime}$ 
radius.  
Thus, the PNe and cluster kinematics are roughly consistent 
in M104, out to $\sim$ 100$^{\prime \prime}$.
The PNe velocity dispersion is slightly 
lower than that of the globular clusters beyond
$\sim$ 120$^{\prime \prime}$, but this difference is not significant
given the uncertainties in the cluster velocity dispersion.

Interestingly, in
the two other galaxies for which we can make a direct comparison
of cluster and PNe kinematics, NGC 1399 and NGC 5128, there is
PNe rotation ($\sim$ 300 km/sec and 100 km/sec respectively),
yet {\it no} rotation is seen in the metal-poor clusters.  
The metal-rich clusters in NGC 5128 have V $\sim$ 40 km/sec inside
6 kpc.
Arnaboldi et al. (1994) attribute the difference in 
kinematics in NGC 1399 to a tidal interaction between it and the 
nearby NGC 1404.  N-body work by Barnes (1996) 
shows that stellar populations 
preserve some kinematic memory after major merger events, and
Barnes speculates that ``... if NGC 5128 is the result of a
major merger, the planetary nebulae and globular clusters may
trace different populations from the original galaxies, with
the former exhibiting a kinematic memory of the disks from whence
they came."

\subsection{The Mean Cluster Metallicity}

Although our cluster spectra are individually too noisy to obtain
useful metallicity estimates, the combination of all 34 spectra into
one higher S/N spectrum does give a good determination of the mean
cluster metallicity.  Bridges \& Hanes (1992) showed that there is
no dependence of mean B$-$V colour on B magnitude or galactocentric
radius; thus, the mean metallicity of our 34 clusters should be
representative of the M104 cluster system as a whole.  
Our individual spectra were shifted using the cross-correlation
results, and then added to give the final spectrum shown in Figure 4.
Ca H\&K, H$\delta$, G-band and H$\gamma$ 
are all clearly seen.

For quantitative analysis, we have used the G-band equivalent width (EW)
to determine [Fe/H]; the G-band is one of Brodie \& Huchra's (1990;
BH hereafter)
6 primary metallicity indicators.  Following their prescription, we 
fit the continuum on two segments between 4284--4300 $\AA$ and
4336-4351 $\AA$, and calculate the feature EW between 4300-4333 $\AA$.  Using
the FIGARO ABLINE routine, we find a central wavelength of 4317 $\AA$ 
and an EW of 4.40 $\AA$.  We then convert this EW into the BH 
feature index I via I=$-$2.5log(1~$-$~EW/$\delta$$\lambda$), yielding
I=0.155.  Finally, we use the BH calibration between
G-band strength and [Fe/H], as determined from calibrating clusters in the 
Milky Way and M31, to estimate [Fe/H].  We find 
[Fe/H]=$-$0.70 $\pm$ 0.30, where the uncertainty is taken from the
scatter in the BH calibration (their Table 7).

There is another BH primary index that falls within our 
bandpass: $\Delta$, which measures the line-blanketing discontinuity
at 4000 $\AA$.  (Note that the CNB feature, between 3810--3910 $\AA$
is also within our bandpass, but we lack a continuum passband
shortwards of this feature).
Unfortunately, our lack of flux calibration (due to the difficulties
of flux-calibrating multi-slit spectra) makes broad metallicity
indices such as $\Delta$ unreliable.  BH list H$+$K (the Calcium
H \& K lines) as one of their ``poorer" calibrators, but it is useful
as a consistency check on our G-band metallicity.  We measure an
EW of 16.0 $\AA$ for H$+$K (feature between 3935--3995 $\AA$, and
continuum fitted between 3920--3935 and 4000--4010 $\AA$),
which translates into [Fe/H] = $-$0.55 $\pm$ 0.4, where again we
have taken the uncertainty from the scatter in the BH calibration.
Though the H$+$K determination has larger scatter, it is consistent
with the G-band value.

%As well, $\Delta$ is much more
%sensitive to atmospheric dispersion than the other indices,
%particularly with multislits where you cannot observe at the 
%parallactic angle.

The mean cluster [Fe/H] agrees well with that estimated previously
by Bridges \& Hanes (1992), who found [Fe/H]= $-$0.8 $\pm$ 0.25
from B$-$V colours of $\sim$ 130 cluster candidates with
0.3 $\leq$ B$-$V $\leq$ 1.3.  It is encouraging to see the agreement
between spectroscopic and photometric metallicities.  As we show in
Table 4, the M104 globular clusters are seen to be considerably
more metal-rich than those of other spirals; Table 4 also shows that
M104 is comparable to E/gE galaxies both in luminosity and mean
cluster [Fe/H]. 

\begin{table}
\caption{Mean [Fe/H] metallicity of globular cluster systems as
a function of parent galaxy luminosity.  This Table has been adapted
from Figure 7 of Secker et al. (1995).  Data for M104 taken from
this paper, for M81 from Perelmuter, Brodie,
\& Huchra (1995), and for other spirals from Harris (1991). 
Globular cluster data for NGC 1399 from
Ostrov et al. (1993), for M87 from Lee \& Geisler (1993), for NGC
3923 from Zepf et al. (1995), for NGC 6166 from Bridges 
et al. (1996), and for remaining E galaxies from Harris (1991).}
\begin{tabular}{lccc} 
\hline\hline
Galaxy & M$_{V_T}$ & Mean Cluster & 
uncertainty \\
 & & [Fe/H] & \\
\hline
 & & & \\
 & E/gE & & \\
 & & & \\
NGC 1399 & -21.1 & -0.90 & 0.20 \\
NGC 3311 & -22.8 & -0.34 & 0.30 \\
NGC 4486 & -22.4 & -0.86 & 0.20 \\
NGC 3923 & -22.05 & -0.55 & 0.20 \\
NGC 4472 & -22.6 & -0.80 & 0.30 \\
NGC 4649 & -22.2 & -1.10 & 0.20 \\
NGC 5128 & -22.0 &  -0.84 & 0.10 \\
NGC 6166 & -23.6 & -1.00 & 0.30  \\
 & & & \\
 & Spirals & & \\
 & & & \\
M104 &  -22.1  &  -0.70 & 0.30 \\
M33 &     -19.2 &  -1.40 &  0.20 \\
MW  &     -21.3 & -1.35 & 0.05 \\
M31 &     -21.7 & -1.21 & 0.05 \\
M81  &    -21.0 &  -1.50 &  0.20 \\
NGC 3031 &  -21.2 &  -1.46 &  0.31 \\
\hline
\end{tabular}
\end{table}

Much of the recent discussion about a possible relationship between
galaxy luminosity and mean cluster metallicity has focussed on
elliptical galaxies.  It is instructive to see if there is any such
relationship for spirals alone.  Figure 5 shows the data for the
6 spirals in Table 4.  Over the small magnitude range 
$-$21 $\leq$ M$_V$ $\leq$ $-$22, there does seem to be a trend, though
this is largely driven by our new data for M104.  By contrast, there
seems to be considerably more scatter about any possible relationship
for E/gE galaxies (cf. Secker et al 1995).  This difference between
spirals and ellipticals would be expected if most ellipticals are
created by mergers.  During a merger a new generation of clusters
of higher metallicity may be created (AZ), and the galaxy luminosity
will change, creating more scatter in both M$_V$ and cluster [Fe/H].
Spirals, however, have presumably not experienced  major merger 
events, and will adhere more closely to any ``primordial" 
M$_V$--[Fe/H] relation.  However, M104 is not a typical spiral:
it is very bright and its disk is a minor component; it is thus not
clear if it should be included in Figure 4.  M33 is also unusual in
the other extreme, since it has very little stellar bulge yet has
managed to produce a significant cluster population (e.g. Bothun 1992).
More metallicities for globular clusters in spiral galaxies are
urgently needed.

\section{Conclusions}

We have obtained spectra for 76 globular 
cluster candidates in M104.  34 of
these objects have been confirmed as M104 globular clusters from 
their spectra and radial velocities; this sample extends out to
$\sim$ 5.5 arcmin in projected radius (14 kpc for 
our adopted distance of D=8.55 Mpc).
Our main conclusions are as follows:

\noindent{\bf (1):~~}The cluster velocity dispersion is 255 (210, 295)
km/sec, after correction for possible rotation of the cluster
system.  This result confirms that M104 is a very massive spiral, as
would be inferred from its luminosity.

\noindent{\bf (2):~~}The Projected Mass Estimator yields a mass of
5.0 (3.5, 6.7) $\times$ 10$^{11}$ M$\odot$ 
for M104, within a projected radius of
5.5 arcmin (14 kpc).  The corresponding mass-to-light ratio is
M/L$_{B_T}$ = 22$^{+7.5}_{-6.5}$ (M/L$_{V_T}$= 16$^{+5.5}_{-5.0}$),
assuming isotropic orbits and a central point mass for M104.
Although there are considerable uncertainties in this
M/L determination, we believe that our quoted value is at the low
end of the allowed range.  Comparing to the M/L$_V$ $\leq$ 4 found
from stellar and HI rotation curves within 200$^{\prime \prime}$, 
our best
estimate is that the M/L ratio of M104 increases with radius, as would be
expected if the galaxy was surrounded by a dark matter halo.

\noindent{\bf (3):~~} There is a marginal detection of rotation in the
M104 globular cluster system at the 
92.5\% confidence level. 
However, many more cluster velocities are required
to conclusively establish rotation.  The kinematics of the clusters
are roughly consistent with preliminary results of Freeman et al.
for the M104 PNe.

\noindent{\bf (4):~~} The mean globular cluster metallicity as determined
from the G-band equivalent width of the composite cluster spectrum is
[Fe/H]= $-$0.70 $\pm$ 0.3, using the calibration of Brodie \& Huchra
(1990).  This metallicity is higher than that of clusters in other
spiral galaxies, but comparable to that of E/gE galaxies of similar
luminosity to M104.

\section*{acknowledgements}

We would like to acknowledge Harvey MacGillivray for doing the COSMOS
scans, and Dave Malin for expediting the delivery ofthe M104 plates
to Edinburgh.  
We thank Ken Freeman and his collaborators for sharing
preliminary results from their M104 PNe data.  We are grateful to
Nial Tanvir and Karl Glazebrook for all of their help with LEXT.
Many thanks go to Tina Bird for her assistance with ROSTAT and
all things 
robust.  We appreciate the useful discussions with Joe Haller
and Scott Tremaine regarding the use of the Projected Mass Estimator.
We also acknowledge the assistance of Peter Bleackley in data 
reduction.  SEZ acknowledges support from NASA through grant 
number HF-1055.01-93A awarded by the Space Telescope Science Institute,
which is operated by the Association of Universities for Research in
Astronomy, Inc., for NASA under contract NASA-26555.  DAH and JJK 
acknowledge NSERC for support through an Operating Grant provided
to DAH.  This research has made use of the NASA/IPAC Extragalactic Database    
(NED)
   which is operated by the Jet Propulsion Laboratory, California Institute   
   of Technology, under contract with the National Aeronautics and Space      
   Administration.

\newpage

\begin{figure*}
\epsfysize 2.5truein
\hfil\epsffile{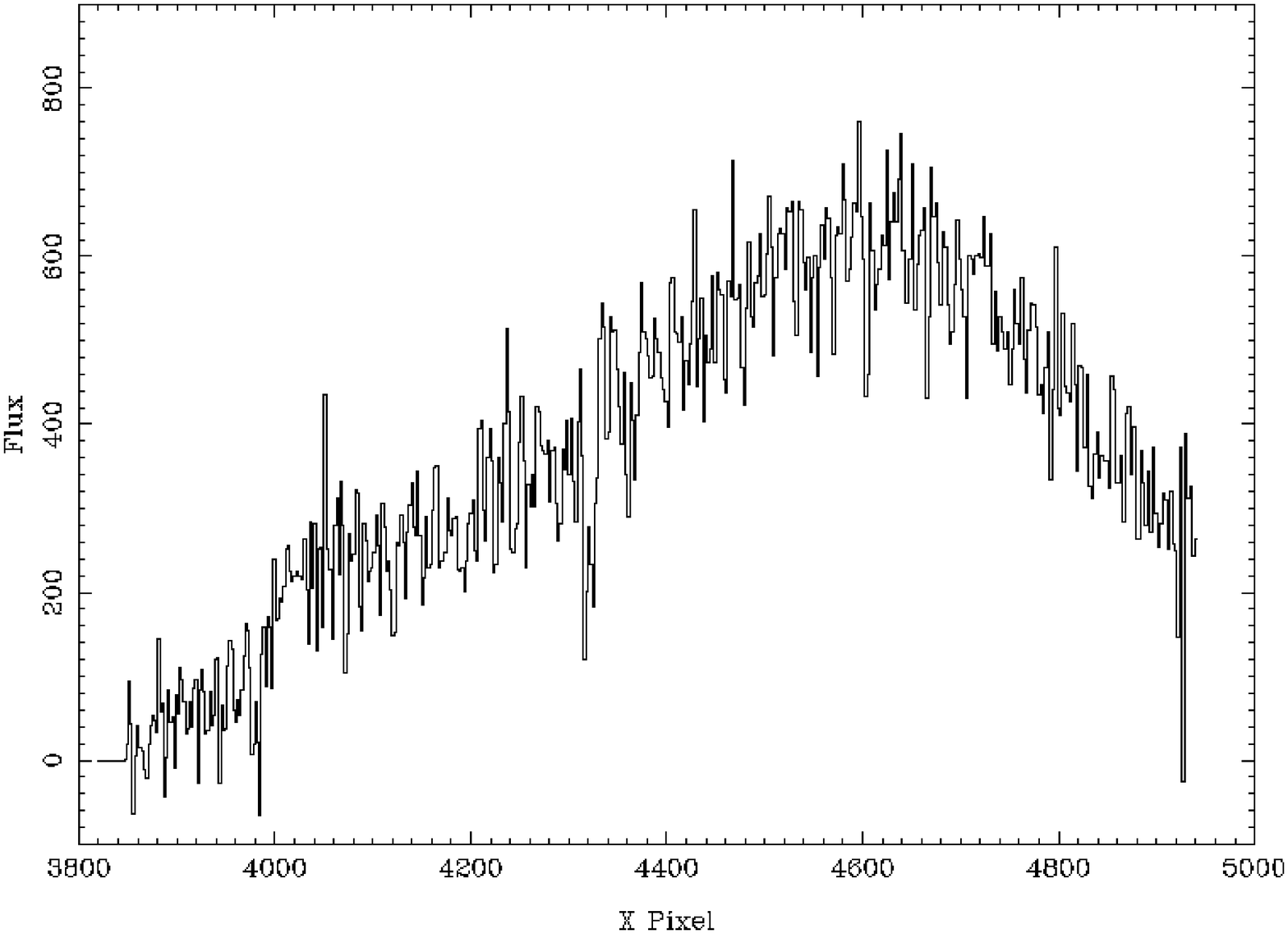}\hfil
\end{figure*}
\begin{figure*}
\epsfysize 2.5truein
\hfil\epsffile{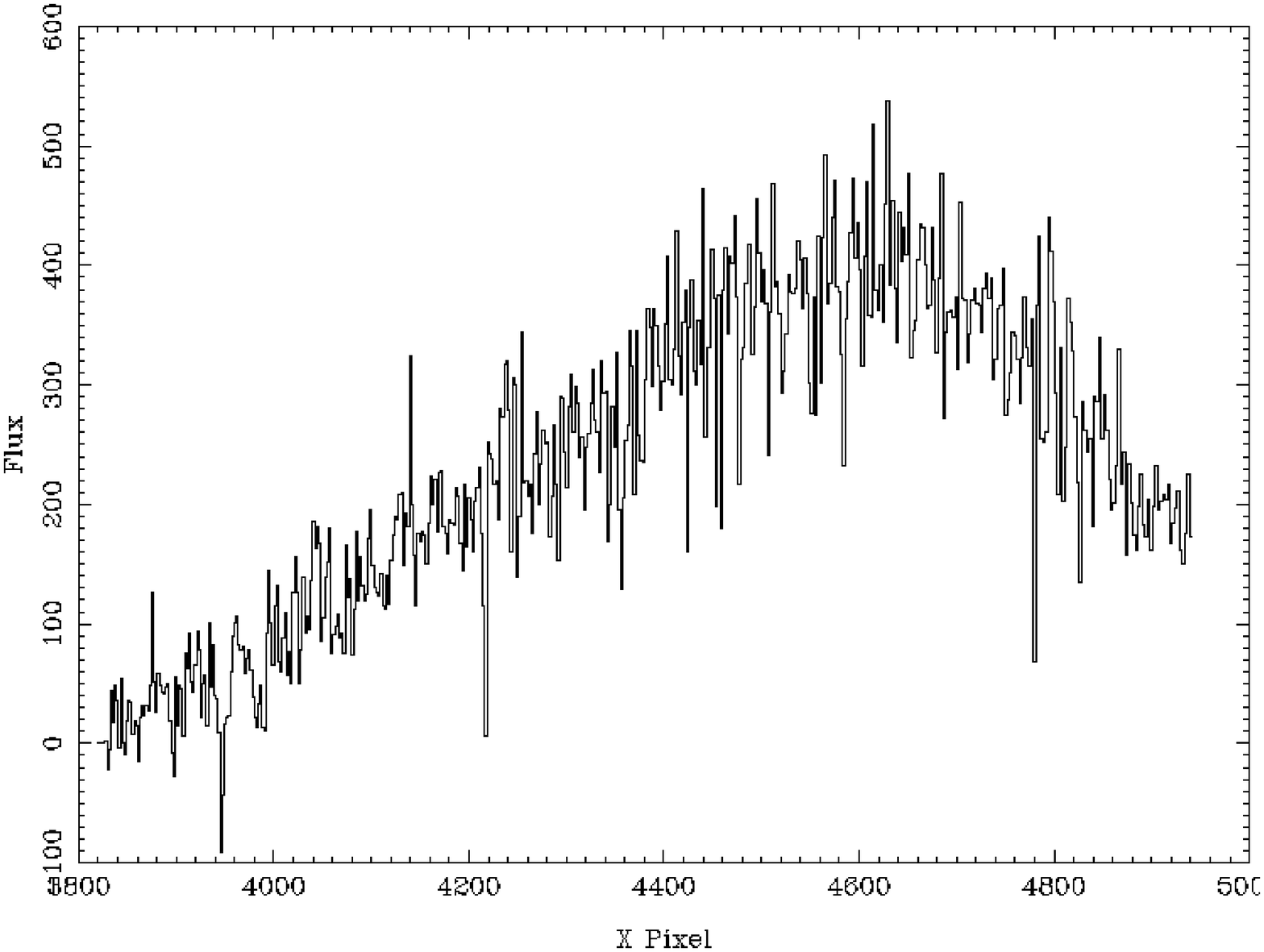}\hfil
\end{figure*}
\begin{figure*}
\epsfysize 2.5truein
\hfil\epsffile{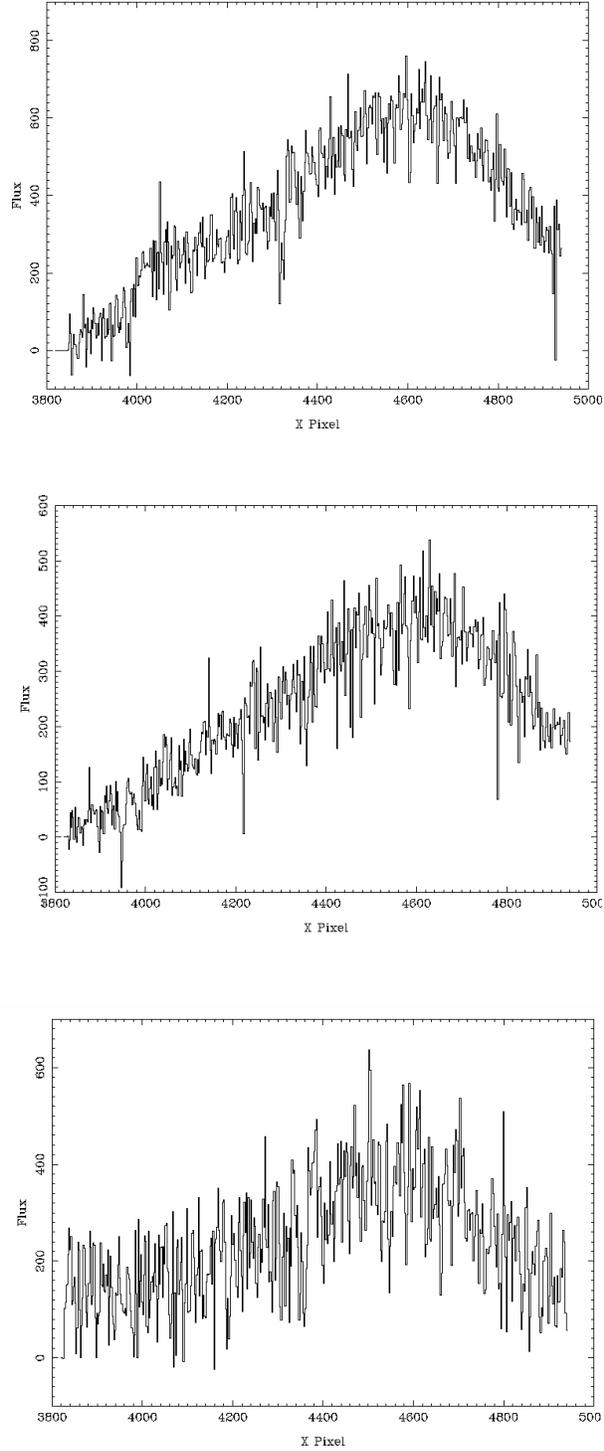}\hfil
\caption{Representative Spectra of M104 Globular Clusters.  The top,
middle and bottom panels show spectra of high (\#1--8), average
(\#2--9), and low (\#1--36) S/N, respectively.  Units along the X
axis are in $\AA$, while the Y axis units are in counts.}
\end{figure*}

\newpage

\begin{figure*}
\epsfysize 4.0truein
\hfil\epsffile{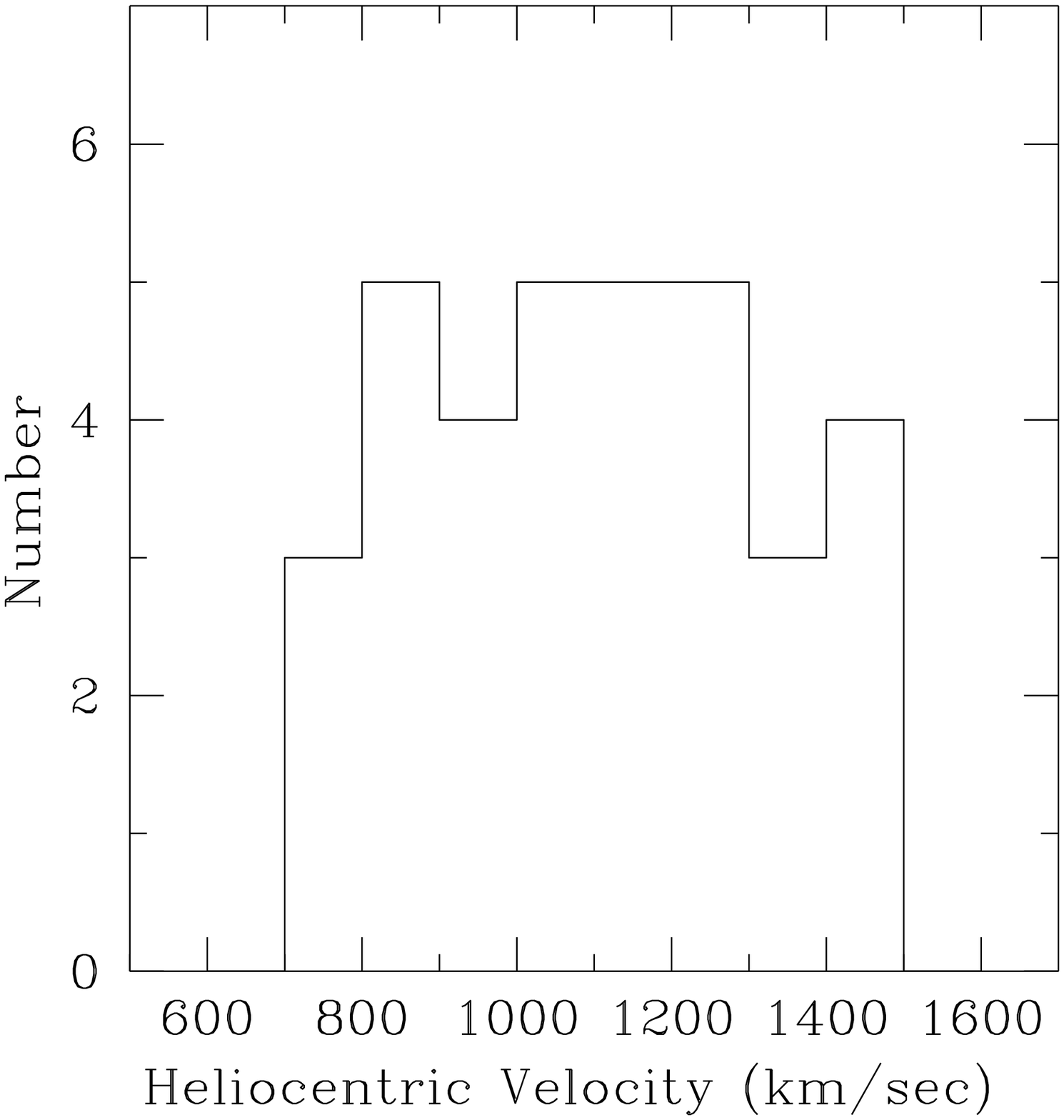}\hfil
\caption{Velocity Histogram of (34) Confirmed M104 Globular Clusters.}
\end{figure*}

\bigskip

\begin{figure*}
\epsfysize 4.0truein
\hfil\epsffile{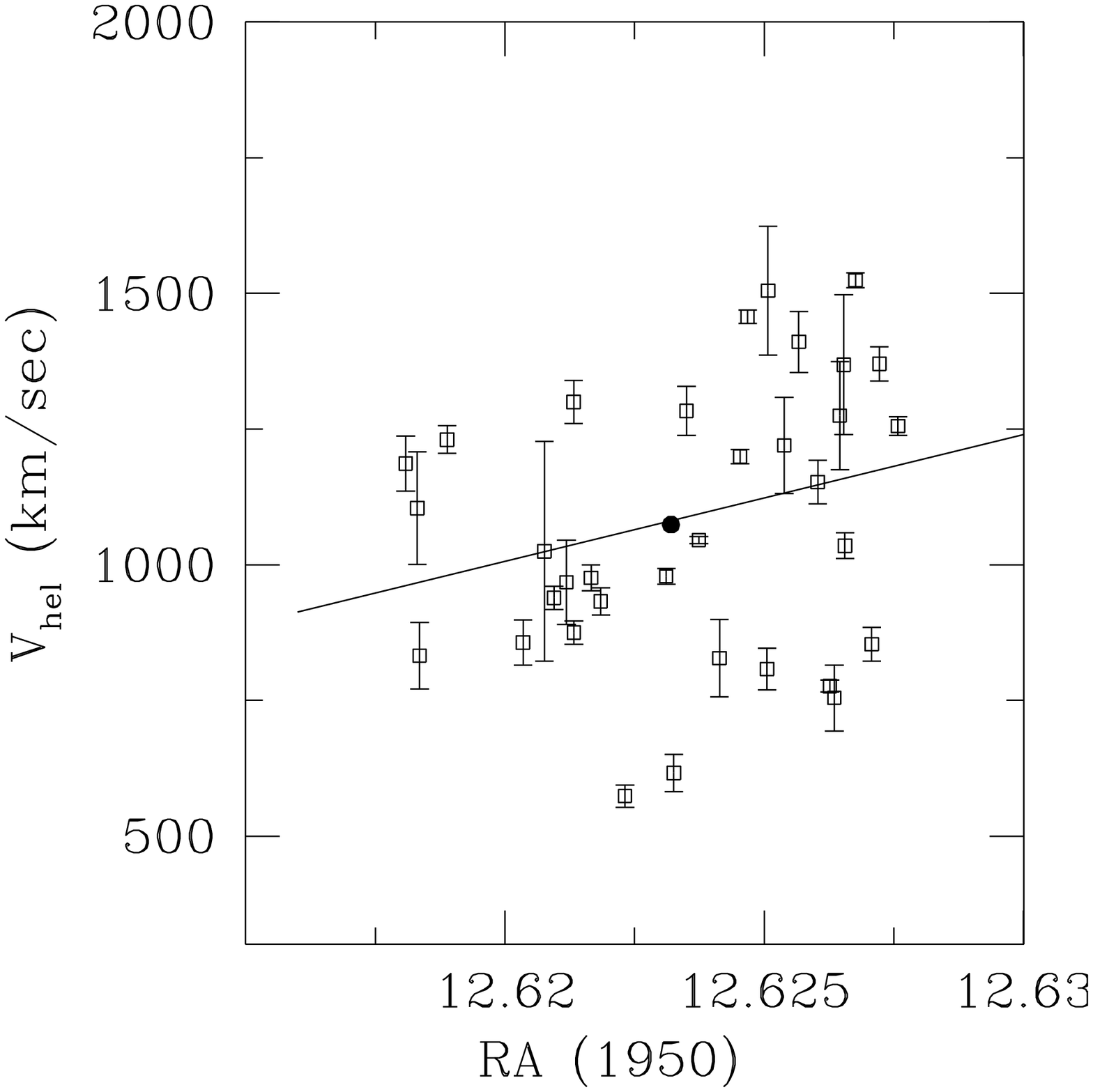}\hfil
\caption{Globular Cluster Velocity vs. Major Axis Radius.  The black
circle denotes the galaxy center.}
\end{figure*}

\newpage

\begin{figure*}
\epsfysize 4.0truein
\hfil\epsffile{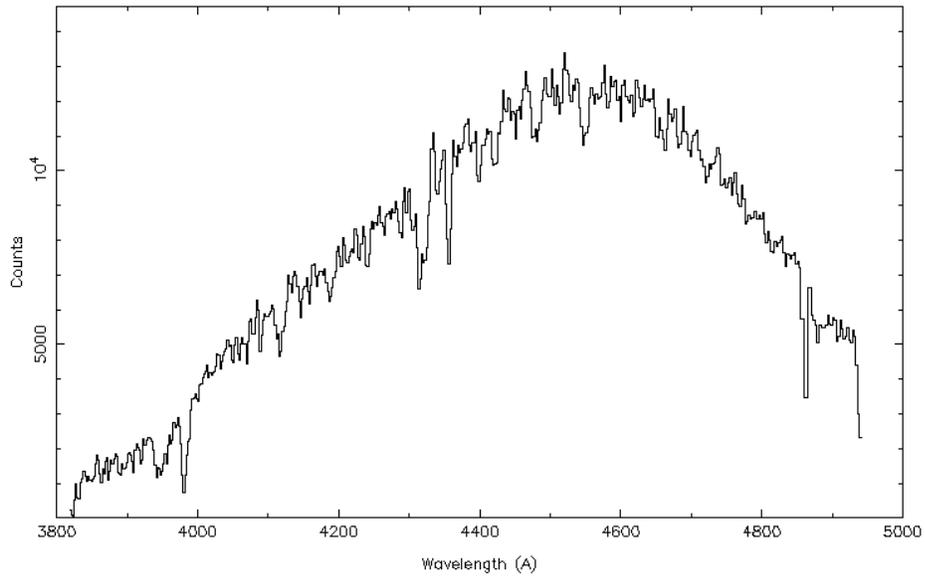}\hfil
\caption{Co-added Spectrum of (34) confirmed M104 globular clusters.}
\end{figure*}

\bigskip

\begin{figure*}
\epsfysize 4.0truein
\hfil\epsffile{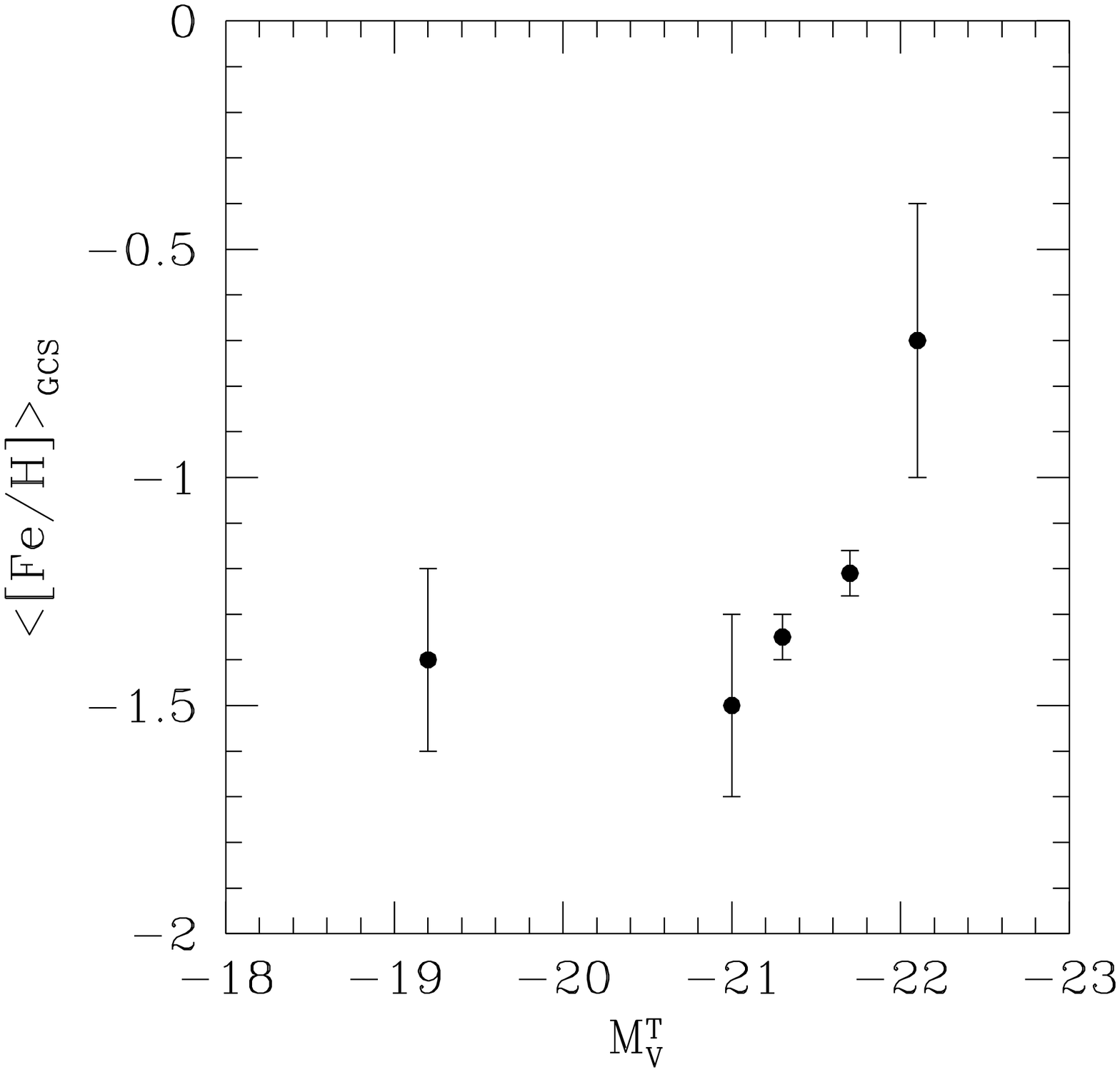}\hfil
\caption{Mean Cluster [Fe/H] vs. Parent Galaxy Luminosity (M$_{V_T}$)
for spiral galaxies.}
\end{figure*}


\newpage


\begin{thebibliography}{}

\bibitem{} Armandroff, T.E., 1989, AJ, 97, 375

\bibitem{} Armandroff, T.E., Zinn, R., 1988, AJ, 96, 92

\bibitem{} Arnaboldi, M., Freeman, K.C., Hui, X., Capaccioli, M.,
Ford, H., 1994, Eso Messenger, 76, 40

\bibitem{} Ashman, K.M., Bird, C.M., Zepf, S.E., 1994, AJ, 108, 2348

\bibitem{} Ashman, K.M., Bird, C.M., 1993, AJ, 106, 2281

\bibitem{} Bahcall, J.N., Tremaine, S., 1981, ApJ, 244, 805 

\bibitem{} Barnes, J.E., 1996, in {\it Formation of the
Galactic Halo ... Inside and Out}, ASP Conference Series
\#92, eds. H. Morrison \& A. Sarajedini, (ASP: San Francisco),
pg. 415

\bibitem{} Beers, T.C., Flynn, K., Gebhardt, K., 1990, AJ, 100, 32

\bibitem{} Bird, C.M., Beers, T.C., 1993, AJ, 105, 1596

\bibitem{} Bothun, G.D., 1992, AJ, 103, 104

\bibitem{} Bridges, T.J., Hanes, D.A., 1992, AJ, 103, 800

\bibitem{} Bridges, T.J., Carter, D., Harris, W.E., Pritchet, C.J.,
1996, MNRAS, in press

\bibitem{} Brodie, J.P., 1993, in {\it The Globular Cluster-Galaxy
Connection}, ASP Conference Series \#48, pg. 483 

\bibitem{} Brodie, J.P., Huchra, J.P., 1990, ApJ, 362, 503 (BH)

\bibitem{} Burkhead, M.S., 1986, AJ, 91, 777

\bibitem{} Burstein, D., Heiles, C., 1984, ApJS, 54, 33

\bibitem{} Ciardullo, R., Jacoby, G.H, Tonry, J.L., 1993, ApJ, 419, 479

\bibitem{} Da Costa, G.S., Armandroff, T.E. 1995, AJ, 109, 2533

\bibitem{} Faber, S.M., Balick, B., Gallagher, J.S., Knapp, G.R.,
1977, ApJ, 214, 383

\bibitem{} Grillmair, C.J., Freeman, K.C., Bicknell, G.V.,
Carter, D., Couch, W.J., Sommer-Larsen, J., Taylor, K., 1994,
ApJ, 422, 9

\bibitem{} Harris, W.E., 1991, ARA\&A, 543

\bibitem{} Harris, W.E., Harris, H.C., Harris, G.L.H.,
1984, AJ, 89, 216

\bibitem{} Harris, H.C., Harris, G.L.H., Hesser, J.E., 1988, in
{\it The Harlow-Shapley Symposium on Globular Cluster Systems
in Galaxies}, pg. 205

\bibitem{} Heisler, J., Tremaine, S., Bahcall, J.N., 1985, ApJ, 298, 8

\bibitem{} Hes, R., Peletier, R.F., 1993, A\&A, 268, 539

%\bibitem{} Hesser, J.E. 1993, in {\it The Globular Cluster-Galaxy
%Connection}, ASP Conference Series \#48, pg. 1

\bibitem{} Huchra, J.P., 1993, in {\it The Globular Cluster-Galaxy
Connection}, ASP Conference Series \#48, pg. 420 

\bibitem{} Hui, X., Ford, H.C., Freeman, K.C., Dopita, M.A., 
1995, ApJ, 449, 592

\bibitem{} Jarvis, B.J., Freeman, K.C. 1985, ApJ, 295, 324

\bibitem{} Jarvis, B.J., Dubath, P., 1988, A\&A, 201, 33

\bibitem{} Kormendy, J., 1988, ApJ, 335, 40

\bibitem{} Kormendy, J., \& Illingworth, G., 1982, ApJ, 245, 460

\bibitem{} Kormendy, J., \& Westpfahl, D.J., 1989, ApJ, 338, 752

\bibitem{} Lee, M.G., Geisler, D., 1993, AJ, 106, 493

\bibitem{} McLachlan, G.J., Basford, K.E., 1988, in
{\it Mixture Models: Inference and Applications to Clustering},
(Marcel Dekker, New York).

\bibitem{} Mould, J.R., Oke, J.B., Nemec, J.M., 1987, AJ, 93, 53

\bibitem{} Mould, J.R., Oke, J.B., de Zeeuw, P.T., Nemec, J.M.,
1990, AJ, 99, 1823

\bibitem{} Ostrov, P., Geisler, D., Forte, J.C., 1993, AJ, 105, 1762

\bibitem{} Perelmuter, J., Brodie, J.P., Huchra, J.P., 1995,
AJ, 110, 620

\bibitem{} Sandage, A., Tammann, G.A., 1981, {\it A Revised
Shapley-Ames Catalog of Bright Galaxies}, Carnegie Institution
of Washington Publication No. 635, Washington, DC

\bibitem{} Schommer, R.A., Christian, C.A., Caldwell, N.,
Bothun, G.D., Huchra, J.A., 1991, AJ, 101, 873

%\bibitem{} Schommer, R.A., 1993, in {\it The Globular Cluster-Galaxy
%Connection}, ASP Conference Series \#48, pg. 458 

\bibitem{} Secker, J., Geisler, D., McLaughlin, D.E., Harris, W.E.,
1995, AJ, 109, 1019

\bibitem{} Schweizer, F., 1978, ApJ, 220, 98

\bibitem{} Wakamatsu, K.I., 1977, PASP, 89, 267

\bibitem{} Zepf, S.E. 1995, in {\it Dark Matter}, eds.
S.S. Holt \& C.L. Bennett, (AIP: New York), pg. 153

\bibitem{} Zepf, S.E., Ashman, K.M., 1993, MNRAS, 264, 611

\bibitem{} Zepf, S.E., Ashman, K.M., Geisler, D., 1995, ApJ, 443, 570

\bibitem{} Zinn, R., 1985, ApJ, 293, 424


\end{thebibliography}
\end{document}